\documentclass[prb, 11pt, superscriptaddress, notitlepage]{revtex4-1}
\usepackage[english]{babel}

\usepackage{subfigure}
\usepackage{siunitx}

\usepackage{natbib}
\usepackage{graphicx}
\usepackage{dcolumn}
\usepackage{amsmath}
\usepackage{amssymb}
\usepackage[dvipsnames]{xcolor}
\usepackage{bm}

\newcommand{\bff}{\mathbf{f}}
\newcommand{\bp}{\mathbf{p}}

\newcommand{\bn}{\mathbf{n}}

\newcommand{\bs}{\mathbf{s}}
\newcommand{\bj}{\mathbf{j}}
\newcommand{\bz}{\hat{\mathbf{z}}}

\newcommand{\bx}{\hat{\mathbf{x}}}
\newcommand{\bsigma}{\boldsymbol{\sigma}}
\newcommand{\bcalA}{\mathbf{A}}
\newcommand{\bcalF}{\mathbf{F}}
\newcommand{\bcalB}{\mathbf{B}}

\newcommand{\taus}{\tau_s}
\newcommand{\tauDP}{\tau_\textnormal{\tiny DP}}

\newcommand{\ls}{l_s}
\newcommand{\lDP}{l_\textnormal{\tiny DP}}
\newcommand{\sigmaD}{\sigma_\mathrm{D}}
\newcommand{\gr}{{g_r^{\uparrow\downarrow}}}
\newcommand\average[1]{{\langle{#1}\rangle}}
\newcommand{\re}{\mathrm{e}}
\newcommand{\rd}{\mathrm{d}}
\newcommand{\ri}{\mathrm{i}}

\begin{document}

\title{Spin Hall magnetoresistance and spin Nernst magnetothermopower: role of the inverse spin galvanic effect}
\author{Sebastian T\"{o}lle}
\affiliation{Universit\"at Augsburg, Institut f\"ur Physik, 86135 Augsburg, Germany}
\author{Michael Dzierzawa}
\affiliation{Universit\"at Augsburg, Institut f\"ur Physik, 86135 Augsburg, Germany}
\author{Ulrich Eckern}
\affiliation{Universit\"at Augsburg, Institut f\"ur Physik, 86135 Augsburg, Germany}
\author{Cosimo Gorini}
\affiliation{Universit\"at Regensburg, Institut f\"ur Theoretische Physik, 93040 Regensburg, Germany}
\date{\today}
\begin{abstract}
In ferromagnet/normal-metal bilayers, the sensitivity of the spin Hall magnetoresistance and the spin Nernst magnetothermopower 
to the boundary conditions at the interface is of central importance.  In general, such boundary conditions 
can be substantially affected by current-induced spin polarizations.  In order to quantify the role of the latter, 
we consider a Rashba two-dimensional electron gas with a ferromagnet attached to one side of the system. 
The geometry of such a system maximizes the effect of current-induced spin polarization on the boundary conditions, 
and the spin Hall magnetoresistance is shown to acquire a non-trivial and asymmetric dependence on the magnetization direction of the ferromagnet.
\end{abstract}


\maketitle

\section{Introduction}
In recent years, the fields of spintronics and spin-calori\-tronics have gained considerable attention \cite{zutic2004, sinova2015, bauer2012}. 
In nonmagnetic materials the most prominent spintronic phenomena are the spin Hall effect, i.e., a transversal spin current due to an applied electrical field \cite{dyakonov1971, kato2004},
and the current-induced spin polarization \cite{ivchenko1978, vasko1979, aronov1989, edelstein1990}. In the literature,
the latter is also referred to as inverse spin galvanic, Rashba-Edelstein, or 
simply Edelstein effect.
The spin-caloritronic counterparts of these electrical effects, exchanging the electrical field with a thermal gradient, are the spin Nernst effect\cite{cheng2008, ma2010, borge2013} and the thermally induced spin polarization \cite{wang2010, dyrdal2013}, respectively.

For a long time only theoretically predicted, the spin Nernst effect was finally observed independently by Sheng et al.\ and Meyer et al.\ in 2016 through the measurement of a spin Nernst signature in the thermopower \cite{sheng_arxiv, meyer_arXiv}. This was accomplished by manipulating the thermally induced spin current in a Pt film by means of the spin transfer torque\cite{slonczewski1996, brataas2000, tserkovnyak2005} induced by attaching an insulating ferromagnet to the metallic film. The resulting thermopower is the thermal analog of the spin Hall magnetoresistance \cite{nakayama2013, chen2013}, and is thus called spin Nernst magnetothermopower \cite{meyer_arXiv}. Experimental investigations of the spin Hall magnetoresistance have so far concentrated on heavy-metal/ferromagnetic-insulator bilayers \cite{nakayama2013, vlietstra2013, hahn2013, meyer2014, goennenwein2015, velez2016}, since thin films of heavy metals like Pt or W exhibit a large spin Hall conductivity \cite{seki2008, liu2011, isasa2015, hao2015}. Theoretical studies based on phenomenological spin diffusion equations qualitatively agree with experimental findings \cite{chen2013}.

In this article we theoretically investigate the spin Hall magnetoresistance and the spin Nernst magnetothermopower in the framework of a two-dimensional electron gas (2DEG) with Rashba spin-orbit coupling. Our approach is based on the generalized Boltzmann equation derived in Ref.~\onlinecite{gorini2010}. Since spin-electric (e.g., spin Hall) and spin-thermoelectric (e.g., spin Nernst) effects in metallic systems are connected by Mott-like formulas \cite{borge2013, toelle2014}, we shall consider both in the following. For Rashba spin-orbit coupling, the inverse spin galvanic effect and the spin Hall effect are related to each other \cite{raimondi2006, gorini2008, raimondi2012, toelle2017}; and, in the presence of a ferromagnetic insulator/2DEG interface, it is apparent that the spin polarization due to the inverse spin galvanic effect influences strongly the spin currents across the interface. 
Therefore it is to be expected that both the spin Hall magnetoresistance and the spin Nernst magnetothermopower in a Rashba 2DEG
are more subtle and complex than the results obtained for heavy-metal/ferromagnet bilayers using a
purely phenomenological approach. 
The goal of this work is to provide a more rigorous derivation of these effects
for a well-defined microscopic model within the framework of the 
quasiclassical kinetic theory.

The paper is organized as follows. In Sec.~\ref{Sec_system} we introduce the system under study and discuss the role of the boundary conditions.
The generalized Boltzmann equation for the Rashba 2DEG is established in Sec.~\ref{Sec_Boltzmann}. Section \ref{Sec_SHE_EE} focuses on the electrical aspects, i.e., the spin Hall effect and the inverse spin galvanic effect in the presence of a ferromagnetic interface. In Sec.~\ref{Sec_chargesector}, we present our results for the spin Hall magnetoresistance and the spin Nernst magnetothermopower. We briefly conclude in Sec.~\ref{Sec_conclusion}.


\section{Statement of the problem}\label{Sec_system}
\begin{figure}[tb]
\includegraphics*[width=0.6\columnwidth]{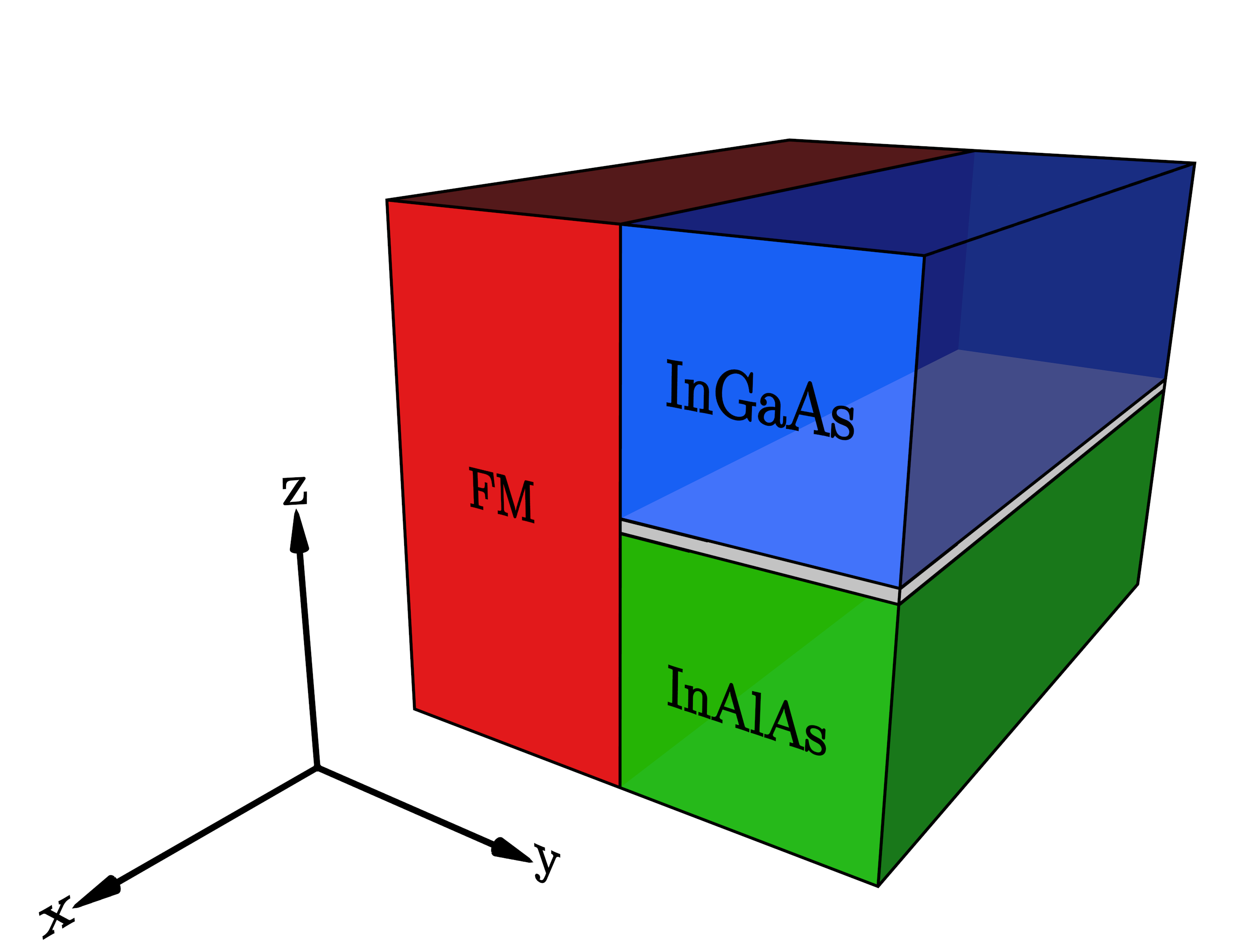}
\caption{Schematic view of a 2DEG, here visualized in grey in a InAlAs/InGaAs heterostructure, in contact with a ferromagnetic insulator (FM). 
The InAlAs/InGaAs heterostructure
is used as an example only: for an experimental realization the materials
need to be chosen so as to minimize upward band bending at the interface with the FM,
which could otherwise deplete the 2DEG in the FM contact region. 
Alternatively, single-crystalline Pt thin films\cite{ryu2016} could be used instead of the semiconductor heterostructure. \label{setup}}
\end{figure} 
A schematic realization of the system under consideration is given in Fig.~\ref{setup}. It consists of a 2DEG in the $x-y$ plane with finite width $L$ in $y$ direction, and an interface to an insulating ferromagnet at $y=0$. By varying the magnetization direction $\bn$ of the ferromagnet it is possible to control the spin current across the interface due to the spin transfer torque. 
More explicitly, the boundary condition for $\bj_y$ (the spin current in $y$ direction), reads
\begin{equation} \label{Eq_FM_boundary}
\bj_y (y=0) = \frac{\gr}{2\pi\hbar N_0} \bn \times \big(\bn \times \bs(y=0)\big) \, ,
\end{equation}
where $\bs$ is the spin density, $N_0 = m/2\pi\hbar^2$ is the density of states per spin and area, and $\gr$ is the real part\footnote{We neglect the imaginary part of the spin mixing 
conductance and its influence on the spin transfer torque since it is usually small compared 
to the real part.} of the spin mixing conductance \cite{tserkovnyak2005}.
In the literature\cite{chen2013, sinova2015} the following simple estimate of the resulting spin Hall magnetoresistance (SMR) due to the boundary condition (\ref{Eq_FM_boundary}) is given: 
assuming that an electrical field $\mathbf{E} = E_x \mathbf{e}_x$ generates a spin polarization $\bs \sim \mathbf{e}_z$, one obtains $\bj_y \sim \bn \times (\bn \times \mathbf{e}_z) $, according to the boundary condition. Due to the inverse spin Hall effect, an additional electrical field $\mathbf{E} \sim \mathbf{e}_y\times \bj_y$ is generated with a magnetization dependence $E_x \sim 1-n_z^2$. For a magnetization within the $y-z$ plane, $\bn = (0,\cos \phi, \sin \phi)$, the resulting SMR signal as function of $\phi$ should therefore be symmetric around $\phi = \pi/2$. The above argumentation is the standard explanation of the SMR observed in thin heavy-metal films deposited on ferromagnetic insulators \cite{nakayama2013, goennenwein2015, velez2016}. However, when in addition an in-plane spin polarization $s^y$ due to the inverse spin galvanic effect is taken into account, it is obvious from Eq.~(\ref{Eq_FM_boundary}) that the resulting SMR signal does not necessarily have this symmetry property. 

The model Hamiltonian for the 2DEG with Rashba spin-orbit interaction reads
\begin{equation} \label{Eq_Hamiltonian}
	H = \frac{p^2}{2m}-\frac{\alpha}{\hbar} (\bsigma \times \bz) \cdot \bp + H_\mathrm{imp}\, ,
\end{equation}
where $\alpha$ is the Rashba coefficient, $\bsigma = (\sigma^x,\sigma^y,\sigma^z)$ is the vector of Pauli matrices, and $H_\mathrm{imp}$ describes a random potential due to nonmagnetic impurities.\footnote{Electron-phonon interaction in the high-temperature limit can be treated analogously since then electron-phonon scattering is essentially elastic. See Refs.\ \onlinecite{toelle2014} and \onlinecite{gorini2015}.}
Spin phenomena related to the presence of impurities are denoted as extrinsic effects, in particular, side-jump, skew-scattering, and Elliott-Yafet relaxation. We focus on the limit where the spin Hall effect is dominated by the Rashba spin-orbit coupling, thus we neglect side-jump and skew-scattering. Nevertheless, we still consider Elliott-Yafet relaxation since the bulk spin hall effect vanishes when only intrinsic contributions are considered in the Rashba system with disorder, see Ref.~\onlinecite{raimondi2012}. 

\section{Generalized Boltzmann equation}\label{Sec_Boltzmann}
We use the kinetic theory employed in Ref.~\onlinecite{gorini2010}, with a generalized Boltzmann equation for the $2\times 2$ distribution function $f = f^0 + \bsigma \cdot \mathbf{f}$,  where $f^0$ is the charge and $\mathbf{f}$ the spin distribution function. In the static case the Boltzmann equation reads
\begin{equation} \label{Eq_Boltzmann}
\frac{\bp}{m} \cdot \tilde{\nabla} f + \frac{1}{2} \left\{ \bcalF \cdot \nabla_{\bp} , f \right\} = I_0 + I_\mathrm{ EY} \, ,
\end{equation}
where $\{ \cdot , \cdot \}$ represents the anticommutator. The covariant spatial derivative and the SU(2) Lorentz force with an electrical field $E_x \bx $ are defined by
\begin{align}
\tilde{\nabla} &= \nabla + \frac{\ri}{\hbar} \left[ \bcalA^a\frac{\sigma^a}{2} , \cdot \right] \, ,  \\
\bcalF &= -e E_x \bx - \frac{\bp}{m} \times \bcalB^a \frac{\sigma^a}{2} \, , \\
B_i^a &= - \frac{1}{2\hbar} \epsilon_{ijk} \epsilon^{abc} A_j^b A_k^c \, ,
\end{align}
where $[\cdot,\cdot]$ is the commutator, and the nonzero components of the SU(2) vector potential are $A_y^x = -A_x^y = 2m\alpha/\hbar$ for Rashba spin-orbit coupling, such that the only nonzero component of the spin-dependent magnetic field $B_i^a$ is $B_z^z = - 4m^2\alpha^2/\hbar^3$. A summation over repeated indices is implied.

The Boltzmann equation, Eq.~(\ref{Eq_Boltzmann}), exhibits three relaxation mechanisms: (i) momentum relaxation, (ii) Elliott-Yafet spin relaxation, and (iii) Dyakonov-Perel spin relaxation.
The collision operators on the r.h.s.\ of Eq.~(\ref{Eq_Boltzmann}) describe momentum relaxation due to impurity scattering ($I_0$) with the momentum relaxation rate $1/\tau$, and Elliott-Yafet spin relaxation ($I_\mathrm{ EY}$) with relaxation rate $1/\taus = (\lambda p /2\hbar)^4/\tau$, where $\lambda$ is the effective Compton wavelength \cite{winklerbook}. 
We refer to Refs.~\onlinecite{raimondi2010, toelle2017}, and \onlinecite{gorini2017} for a more detailed discussion of $I_\mathrm{ EY}$. The Dyakonov-Perel relaxation rate due to Rashba spin-orbit coupling is given by $1/\tauDP = (2m\alpha/\hbar^2)^2 D$ with the diffusion constant $D=v_F^2\tau/2$, where $v_F$ is the Fermi velocity.\footnote{Here, we consider the dirty limit. For a more general discussion of $\tauDP$ see Refs.~\onlinecite{toelle2014} and \onlinecite{raimondi2006}.} The length scales associated with $\tauDP$ and $\taus$ are the Dyakonov-Perel and Elliott-Yafet spin diffusion lengths $\lDP = \sqrt{D\tauDP}$ and $\ls = \sqrt{D\taus}$, respectively. In the following we consider the experimentally relevant situation $\taus > \tauDP \gg \tau$ \cite{sanchez2013}.

In order to set the stage we define the relevant physical quantities as follows:
\begin{align}
	j_x &= -2e\int \frac{\rd^2 p}{(2\pi\hbar)^2} \frac{p_x}{m} f^0  \, , \\
	j_i^a &= \int \frac{\rd^2 p}{(2\pi\hbar)^2} \frac{p_i}{m} f^a  \, , \\
	\bs &= \int \frac{\rd^2 p}{(2\pi\hbar)^2} \mathbf{f} \, ,
\end{align}
where $j_x$ is the charge current in $x$ direction with $e = |e|$, $j_i^a$ is the $a$-polarized spin current flowing in $i$ direction, and $\bs$ is the spin density.

\section{Linear response in the spin sector}\label{Sec_SHE_EE}
In this section we shall discuss the spin Hall effect and the inverse spin galvanic effect due to an electrical field applied along the $x$ direction. 
We assume the system to be homogeneous in $x$ direction but inhomogeneous in $y$ direction due to the presence of boundaries. 
We consider the spin sector of the (static) Boltzmann equation and derive coupled diffusion equations for the spin polarization and the spin current as presented in detail 
in App.~\ref{App_spin_sector}. 
For a magnetization $\bn = (0,\cos \phi, \sin \phi)$ the boundary condition (\ref{Eq_FM_boundary}) for the $x$ component of $\bs$ and $\bj_y$ is decoupled from the $y$ and $z$ components. Therefore, it is possible to restrict ourselves to the $y$ and $z$ components of the spin current for which we obtain
\begin{align}
\left( 2- \ls^2 \nabla_y^2 \right) j_y^y &= \frac{\ls^2+\lDP^2}{\lDP} \nabla_y j_y^z \, \label{Eq_jyyDGL} ,\\
\left( 1+\frac{\taus}{\tauDP}- \lDP^2 \nabla_y^2 \right) j_y^z &= - \frac{\ls^2+\lDP^2}{\lDP} \nabla_y j_y^y + 
\frac{\hbar\sigmaD}{2e \epsilon_F \tauDP}  E_x \, , \label{Eq_jyzDGL}
\end{align}
where $\epsilon_F$ is the Fermi energy and $\sigmaD=2 e^2 N_0 D$ the Drude conductivity.
The spin densities $s^y$ and $s^z$ can be expressed in terms of the spin currents,
\begin{align}
s^y ={}& -  \taus\nabla_y j_y^y - \frac{\taus}{\lDP}j_y^z  + \frac{\hbar\sigmaD}{4 e \epsilon_F \lDP}  E_x \, , \label{Eq_syDGL} \\
s^z ={}& -\tauDP \nabla_y j_y^z + \frac{\tauDP}{\lDP} j_y^y \, , \label{Eq_szDGL}
\end{align}
such that it is straightforward to obtain the spin densities once Eqs.\ (\ref{Eq_jyyDGL}) and (\ref{Eq_jyzDGL}) are solved. 
In the homogeneous case the solutions of the spin diffusion equations are $j_y^y = s^z = 0$, and
\begin{align}
j_y^z = j^z_0 ={}& \frac{\hbar\sigmaD}{2e\epsilon_F(\tauDP+\taus)} E_x \, , \label{Eq_jyz0} \\
s^y = s^y_0 ={}& -\frac{\taus-\tauDP}{2\lDP}  j_0^z \, . \label{Eq_sy0}
\end{align}
The corresponding transport coefficients $\sigma_0^\mathrm{sH}$ and $P_0^{\mathrm{E}}$ are defined through $j^z_0 = \sigma_0^\mathrm{sH} E_x$ and
$s^y_0 = P_0^{\mathrm{E}} E_x$, respectively. From Eqs. (\ref{Eq_jyz0}) and (\ref{Eq_sy0}) it follows that in the limit $\taus \to \infty$ there is no spin Hall 
effect, while in the case $\taus = \tauDP$ the inverse spin galvanic effect vanishes. The latter is no longer the case when side-jump or skew scattering are included \cite{gorini2017}.

Next, we shall discuss the influence of the boundary conditions. First, we analyze the spatial profile of the spin polarization 
and the spin currents, and second we determine spatial averages of $j_y^z$ and $s^y$ as function of the magnetization direction.

\subsection{Spatial profile} \label{Sec_spatial_profile}
The coupled differential equations (\ref{Eq_jyyDGL})--(\ref{Eq_szDGL}) sup\-ple\-ment\-ed by appropriate boundary conditions can be solved both analytically, 
see App.\ \ref{App_spin_sector}, and numerically. 
First, we consider symmetric boundary conditions with 
$\bj_y(0)=\bj_y(L)=0$, corresponding to an isolated stripe of width $L$. 
The vanishing of the normal component of the spin current can be justified from the Boltzmann equation when assuming spin-conserving scattering \cite{schwab2006}.
Second, we consider an asymmetric set-up, with $\bj_y(L)=0$ and $\bj_y(0)$ given in Eq.~(\ref{Eq_FM_boundary}),
corresponding to a ferromagnetic insulator with magnetization direction $\bn$ attached to the ``left'' side ($y=0$) of the stripe.
Obviously, symmetric boundary conditions are recovered by setting $\gr = 0$.
In two dimensions, $\gr$ has the dimension of an inverse length. 

\begin{figure}[tb]
\subfigure[]{\includegraphics[width=0.49\columnwidth]{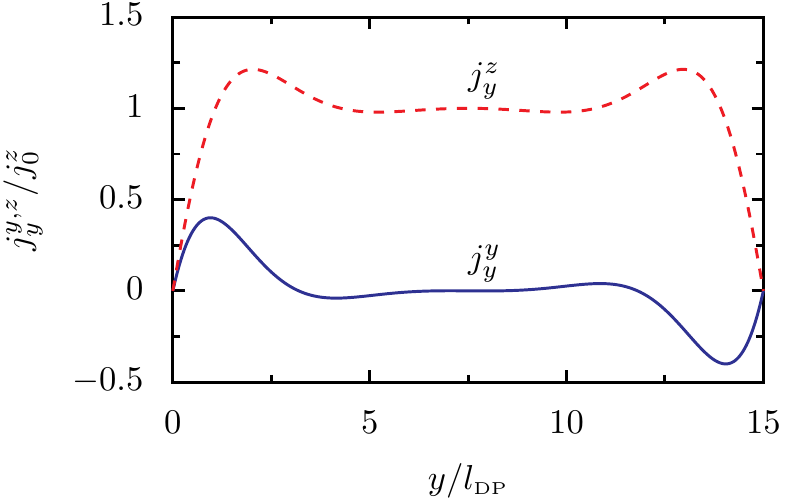}}
\subfigure[]{\includegraphics[width=0.49\columnwidth]{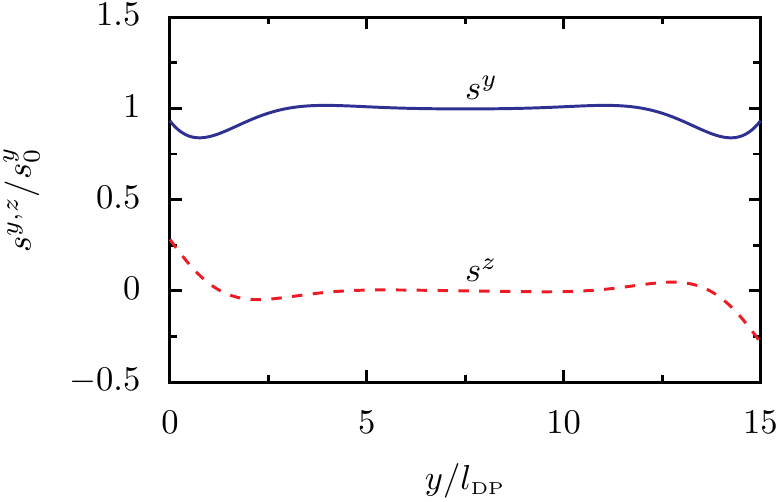}}
\caption{{Spatial profile of the spin currents, (a), and the spin polarizations, (b), for symmetric boundary conditions ($\gr = 0$); $L/\lDP = 15$, $\taus/\tauDP=10$.} \label{Fig_j_sH}}
\end{figure}

Figure \ref{Fig_j_sH} shows the spatial profile of the spin currents and the spin polarizations for symmetric boundary conditions. 
From panel (a) it is apparent that the spin currents exhibit the symmetry $j_y^y(y)=-j_y^y(L-y)$ and $j_y^z(y)=j_y^z(L-y)$, which is consistent with Eqs.~(\ref{Eq_jyyDGL}) and (\ref{Eq_jyzDGL}). Similarly, according to  
Eqs.\ (\ref{Eq_syDGL}) and (\ref{Eq_szDGL}), $s^y(y)=s^y(L-y)$ and $s^z(y)=-s^z(L-y)$, see panel (b). The influence of the boundaries 
is restricted to a range of $\sim 3 \, \lDP$, and thus for larger system sizes it is justified to solve the diffusion equations for a 
semi-infinite system, see App.\ \ref{App_semi}. We obtain:
\begin{align}
j_y^y &= \frac{j_0^z}{2+\ls^2|q|^2} \frac{\lDP |q|^2}{q_{+}} \left( 1+ \frac{\taus}{\tauDP} \right)  \re^{-q_{-} y} \sin(q_{+} y) \, , \\
j_y^z &= j_0^z - \frac{j_0^z }{2+\ls^2|q|^2} \re^{-q_{-} y} \bigg[\!\! \left( 2 + \ls^2|q|^2 \right) \! \cos(q_{+} y) \! + 
\! \frac{q_{-}}{q_{+}} \! \left( 2 - \ls^2|q|^2 \right) \! \sin(q_{+} y) \bigg]  \, ,
\end{align}
where
\begin{equation} \label{Eq_q}
q_\pm = \frac{1}{2\lDP} \sqrt{\sqrt{8+8\frac{\tauDP}{\taus}} \pm \left( 1- \frac{\tauDP}{\taus} \right)} 
\end{equation}
and $|q|^2=q_+^2 + q_-^2$. The symmetrized analytical result deviates by less than $10^{-5}$ from the numerical data shown in Fig.\ \ref{Fig_j_sH}, and even for $L \approx 5\lDP$ analytical and numerical results are still in fair agreement.

\begin{figure}[tb]
\subfigure[]{\includegraphics[width=0.49\columnwidth]{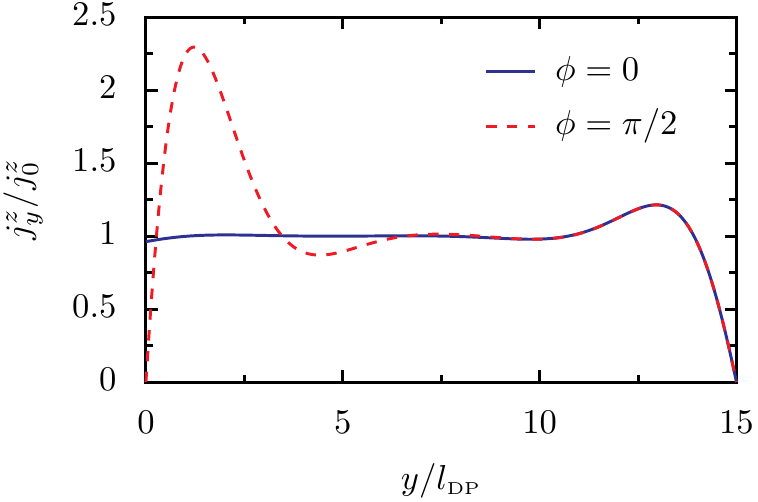}} 
\subfigure[]{\includegraphics[width=0.49\columnwidth]{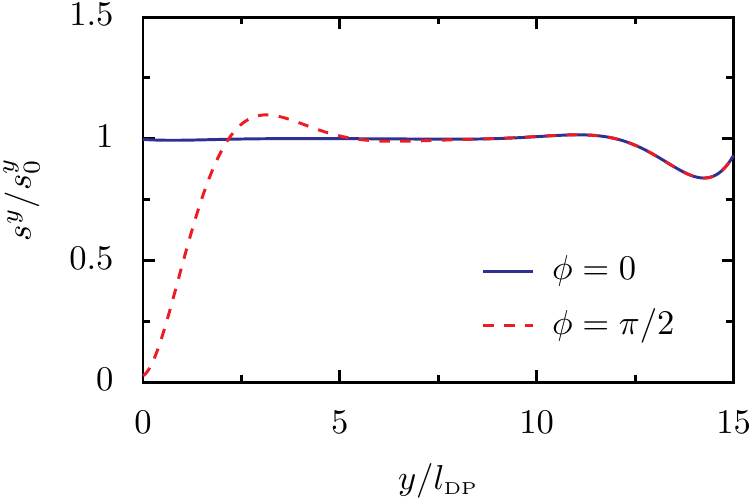}}
\caption{{Spatial profile of the spin current $j_y^z$, (a), and the spin polarization $s^y$, (b), for asymmetric boundary conditions with
$\gr\alpha\tauDP/\hbar=10$ and $\phi = 0, \pi/2$. 
The parameters $L/\lDP$ and $\taus/\tauDP$ are the same as in Fig.\ \ref{Fig_j_sH}.} \label{Fig_jyz}}
\end{figure}

In the case of asymmetric boundary conditions, see Eq.\ (\ref{Eq_FM_boundary}), 
we assume that $\bn$ lies within the $y-z$ plane and is parametrized by $\bn = (0,\cos\phi,\sin\phi)$. Figure \ref{Fig_jyz} shows the spatial 
profile of the spin current $j_y^z$ and the spin polarization $s^y$ for two orientations of the ferromagnetic polarization, $\phi = 0$ and $\phi = \pi/2$. 
A remarkable feature is the hump of $j_y^z$ close to the left boundary for $\phi = \pi/2$. Although the spin current vanishes at the interface, 
the spin current averaged over the whole system can thus be enhanced due to this hump compared to the average spin current in the $\phi = 0$ case. 
The implications of this observation will be discussed in the subsequent section.

\subsection{Spatial averages} \label{Sec_SHE_EE_subResults}
In this subsection, we consider spatial averages of the spin polarization $s^y$ and the spin current $j_y^z$, 
which allows to define an averaged spin Hall conductivity and polarization coefficient, respectively; and we focus on their dependence on the polarization angle $\phi$
of the attached ferromagnet. 
For a stripe of width $L$, the spatial averages of $s^y$ and $j_y^z$, and the corresponding averaged transport coefficients $P_\mathrm{sE}$ and $\sigma_\mathrm{sE}$,
are defined as
\begin{align} \label{Eq_averages}
\langle s^y\rangle = \frac{1}{L}\int_0^L dy\, s^y = P_\mathrm{sE} E_x\\
\langle j_y^z\rangle = \frac{1}{L}\int_0^L dy\, j_y^z = \sigma_\mathrm{sE} E_x \, .
\end{align}
The subscript ``$\mathrm{sE}$'' indicates the linear response of the spin (current or polarization) to an applied electrical field (in contrast
to the linear spin response to a temperature gradient labeled by ``$\mathrm{sT}$'' that will be discussed in Sec.\ \ref{Sec_chargesector}). 

\begin{figure}[tb]
\subfigure[]{\includegraphics[width=0.49\columnwidth]{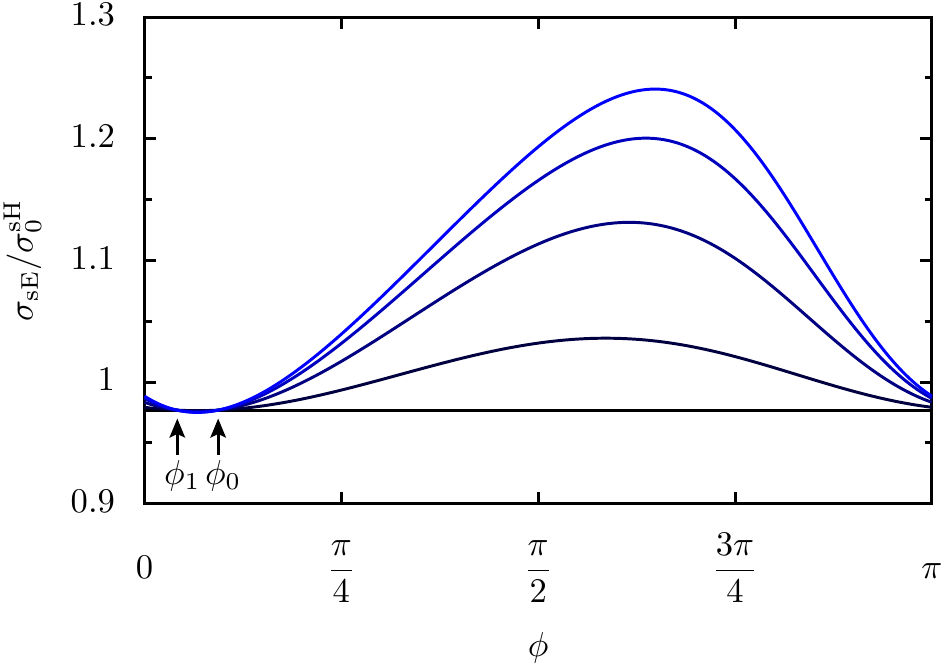}}
\subfigure[]{\includegraphics[width=0.49\columnwidth]{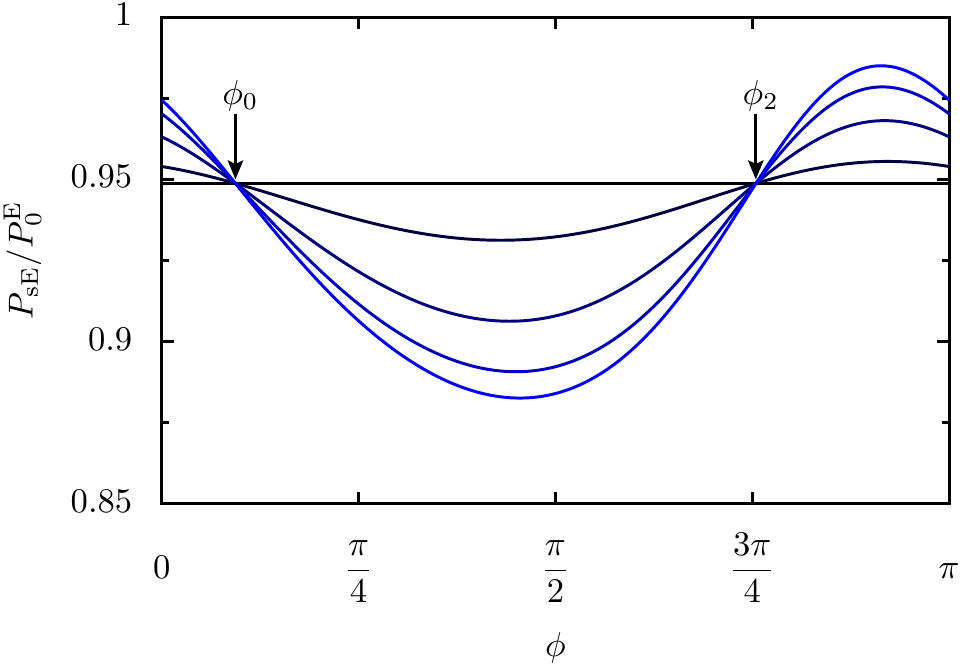}}
\caption{Averaged spin Hall conductivity, (a), and polarization coefficient, (b), versus $\phi$, normalized by their respective bulk values, for $\taus/\tauDP = 10$, $L/\lDP = 10$, and $\gr \alpha \tauDP /\hbar = 0, 0.2, 0.5, 2, 100$ from black to blue. \label{Fig_angles_vargr}}
\end{figure}

Figure \ref{Fig_angles_vargr} shows the averaged spin Hall conductivity, panel (a), and the averaged polarization coefficient, panel (b), 
normalized to their respective bulk values versus the magnetization angle $\phi$ for $L/\lDP = 10$ and various values of the spin mixing conductance $\gr$. 
While the averaged spin Hall conductivity, (a), increases with increasing $\gr$ for nearly all angles $\phi$, with the strongest response in
the range $\pi/2 \lesssim \phi \lesssim 3\pi/4$, the polarization coefficient, (b), can be enhanced or reduced, depending on $\phi$.

In the limit $L\gg\lDP$ it is straightforward to calculate analytically the ferromagnetic contribution of the spin current, defined as 
\begin{equation} \label{Eq_defFM}
\Delta j_y^z = j_y^z - j_y^z(\gr=0) \, ,
\end{equation}
see Eq.\ (\ref{Eq_jyzFM}) in App.\ \ref{App_semi}. Performing the spatial average yields the ferromagnetic contribution to the spin Hall conductivity:
\begin{equation} \label{Eq_sigmaFM}
\frac{\Delta\sigma_\mathrm{sE}}{\sigma_0^\mathrm{sH}} = \frac{2\left(1+\taus/\tauDP\right) j_y^y(0) + 4\lDP q_{-} j_y^z(0)}{ L \lDP |q|^2 \left(2+\ls^2|q|^2\right)j_0^z}  \, .
\end{equation}
Obviously, $\Delta \sigma_\mathrm{sE}$ is fully determined by the boundary values of the spin current, $j_y^y(0)$ and $j_y^z(0)$, which can be controlled by the magnetization angle $\phi$, see Eq.\ (\ref{Eq_FM_boundary}). For $\phi = 0 $ the spin current $j_y^y(0)$ vanishes, and $j_y^z(0) \sim s^z(0)$, while for $\phi=\pi/2$ the spin current $j_y^z(0)$ vanishes, and $j_y^y(0) \sim s^y(0)$. This explains why in the limit $\taus/\tauDP \gg 1$ the averaged spin Hall conductivity $\sigma_\mathrm{sE}$ is enhanced for $\phi \approx \pi/2$ compared to $\phi \approx 0 $ as observed in Fig.\ \ref{Fig_angles_vargr} (a). The above argumentation crucially depends on the existence of a nonvanishing in-plane spin polarization, i.e., the inverse spin galvanic effect.

Remarkably, for the magnetization angle $\phi_0 \approx 0.294$, both $\sigma_\mathrm{sE}$ and $P_\mathrm{sE}$ are independent of $\gr$. 
This is due to the fact, that for this particular angle the spin polarization at the interface, $\bs(\gr=0,y=0)$, is proportional to
the magnetization direction $\bn$, and thus, according to Eq. (\ref{Eq_FM_boundary}), the spin current $\bj_y (0)$ vanishes, independently of $\gr$.
In the limit $L \gg \lDP$, it is possible to calculate $\phi_0$ explicitly, see App.\ \ref{App_semi}, with the result
\begin{equation} \label{Eq_phi0}
\tan \phi_0 =  \frac{4 \tauDP \lDP q_{-} }{\taus + \tauDP(1-\lDP^2 |q|^2)} \, ,
\end{equation}
which yields $\phi_0 \approx 0.2934$, very close to the numerical result for $L=10\lDP$. In addition, $\sigma_\mathrm{sE}$ and $P_\mathrm{sE}$ are also independent of $\gr$ for $\phi_1 \approx 0.131 $ and $\phi_2 \approx 2.37 $, respectively, as indicated by the arrows in Fig.\ \ref{Fig_angles_vargr}. According to Eq.\ (\ref{Eq_sigmaFM}), $\Delta \sigma_\mathrm{sE}$ vanishes if the condition
\begin{equation}
\frac{j_y^y(0)}{j_y^z(0)} = - \frac{2 \tauDP \lDP q_-}{\tauDP+\taus}
\end{equation} 
is fulfilled. On the other hand, due to the boundary condition, Eq.\ (\ref{Eq_FM_boundary}), it follows that $\bj_y(0) \sim (0,-\sin\phi,\cos\phi)$ which yields 
\begin{equation} \label{Eq_phi1}
\tan \phi_1 = \frac{2 \tauDP \lDP q_-}{\tauDP+\taus} \, .
\end{equation}
A similar kind of reasoning for the $\gr$-dependent part of $P_\mathrm{sE}$ leads to
\begin{equation} \label{Eq_phi2}
\tan\phi_2 = -\frac{2q_-}{\lDP|q|^2} \, .
\end{equation} 
Although Eqs.\ (\ref{Eq_phi1}) and (\ref{Eq_phi2}) are strictly valid only in the limit $L \gg \lDP$, the values for $\phi_1$ and $\phi_2$ obtained from Eqs.\ (\ref{Eq_phi1}) and (\ref{Eq_phi2}) are very close to the numerical results for a system of size $L = 10\, \lDP$.

\begin{figure}[tb]
\subfigure[]{\includegraphics[width=0.49\columnwidth]{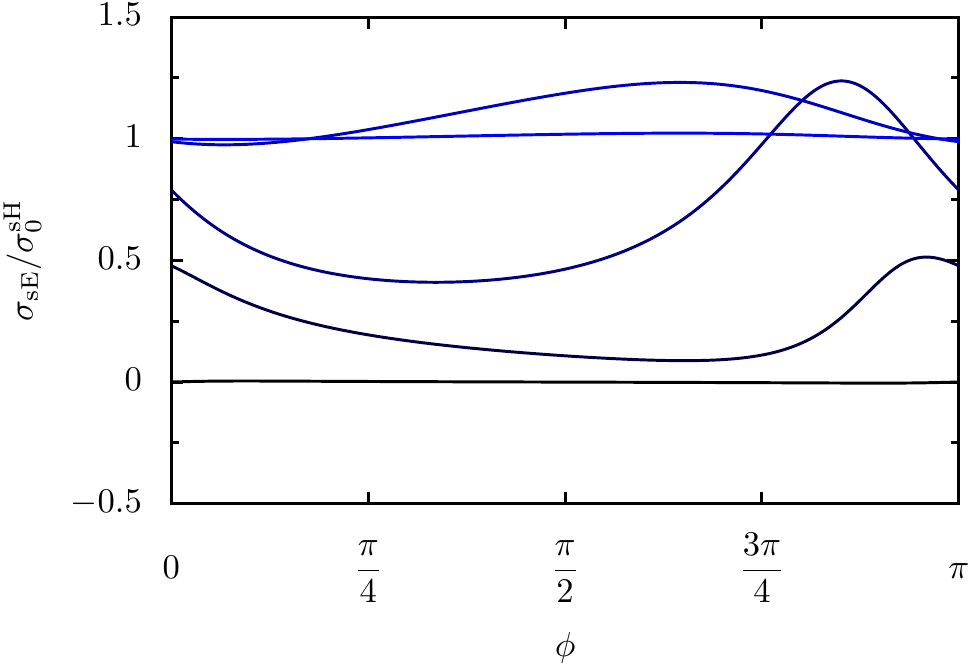}}
\subfigure[]{\includegraphics[width=0.49\columnwidth]{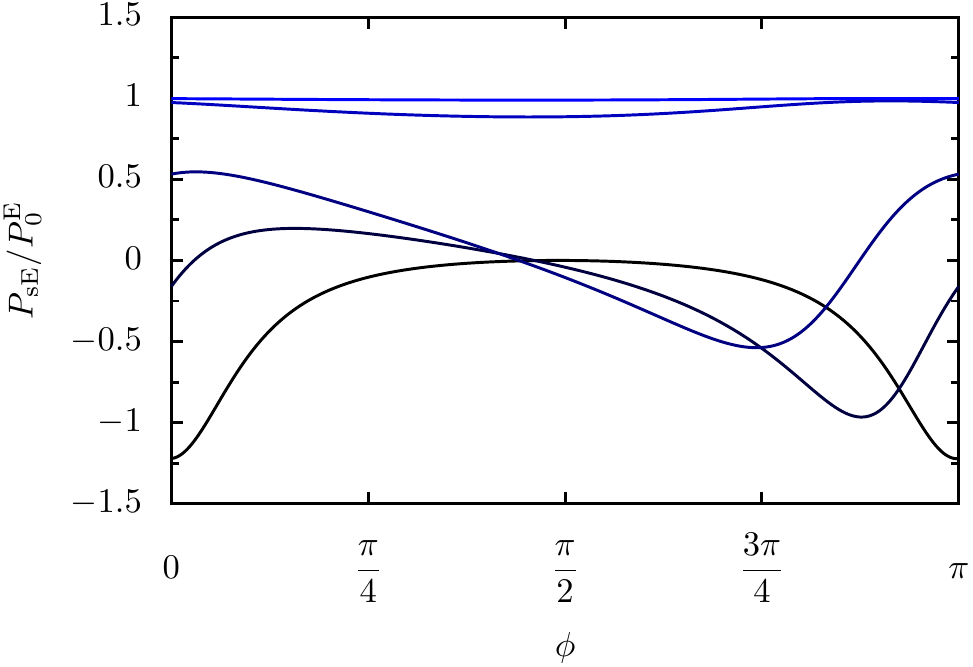}}
\caption{Averaged spin Hall conductivity, (a), and polarization coefficient, (b), versus $\phi$, normalized by their respective bulk values, for $\taus/\tauDP = 10$, $\gr \alpha \tauDP /\hbar = 10$, and $L/\lDP = 0.01, 0.5, 1, 10, 100$ from black to blue. \label{Fig_angles_varL}}
\end{figure}

The averaged spin Hall conductivity, (a), and polarization coefficient, (b), are displayed in Fig.\ \ref{Fig_angles_varL} for fixed spin mixing conductance $\gr\alpha\tauDP/\hbar = 10$ and several values of $L$. Clearly, for very narrow systems, $\sigma_\mathrm{sE}$ has
to go to zero due to the vanishing spin current at the right boundary. In contrast, for very wide systems it has to approach the bulk value
$\sigma_{0}^\mathrm{sH}$ since the influence of the boundary conditions becomes negligible. In between, $\sigma_\mathrm{sE}$ depends nontrivially on the magnetization angle $\phi$. 
The averaged polarization coefficient $P_\mathrm{sE}$ also approaches its bulk value for $L\gg \lDP$. However, in contrast to $\sigma_\mathrm{sE}$,
it does not vanish for very narrow systems, but converges to 
\begin{equation}\label{Eq_smallL}
\frac{P_\mathrm{sE}}{P_0^{\mathrm{E}}} = -\frac{\tauDP(\tauDP + \taus)}{(\taus - \tauDP)(\tauDP+ \taus \tan^2 \phi)}  \, ,
\end{equation}
which is symmetric around $\phi=\pi/2$. Equation (\ref{Eq_smallL}) is obtained by assuming that spin densities and spin currents depend only linearly on $y$, which is justified for $L \ll \lDP$.

\section{Linear response in the charge sector} \label{Sec_chargesector}
In the previous section, we have considered the spin polarization and spin currents in response to an applied electrical field, and pointed out how they can be modulated by changing the magnetization angle of the attached ferromagnet. Since spin signatures (polarization and currents) are notoriously difficult to detect directly in experiment, we consider now the associated signals in the charge current. Furthermore, we extend our analysis by including also thermal effects, i.e., contributions due to a temperature gradient. In particular, we focus on the SMR and the spin Nernst magnetothermopower (SNMTP), i.e., the fingerprint of the magnetization dependent spin Hall and spin Nernst effect in the conductivity and the thermopower, respectively. 

The momentum integrated charge sector of the Boltzmann equation yields the following expression for the width-averaged charge current in linear response to an electrical field $E_x$ and a thermal gradient $\nabla_x T$ (see also Ref.~\onlinecite{toelle2017}):
\begin{equation}
\average{j_x} = \sigmaD E_x - \sigmaD S_0 \nabla_x T  - 2e \frac{\alpha}{\hbar} \frac{\tau}{\lDP}  \left(\average{ j_y^z} - \frac{\lDP}{\taus} \average{s^y} \right) \, . \label{Eq_jx2}
\end{equation}
Here, $S_0 = - \pi^2 k_B^2 T / (3e\epsilon_F)$ is the Seebeck coefficient of a free electron gas, and $\sigmaD$ is the Drude conductivity.
The corresponding expressions for the spin current and the spin polarization are:\cite{toelle2014}
\begin{align}
\average{j_y^z} &= \sigma_\mathrm{ sE} E_x + \sigma_{\rm sT} \nabla_x T \,  \label{Eq_jyzgeneral},\\
\average{s^y} &= P_\mathrm{ sE} E_x + P_\mathrm{ sT} \nabla_x T \, , \label{Eq_sygeneral}
\end{align}
respectively, where the direct spin Nernst and the direct thermal polarization coefficients are given by\cite{toelle2014}
\begin{align}
\sigma_\mathrm{ sT} \ &=  -S_0 \epsilon_F {\sigma'}_\mathrm{ sE}(\epsilon_F) \, , \label{Eq_DefThermal1} \\
P_\mathrm{ sT} \ &= -S_0 \epsilon_F {P'}_\mathrm{ sE}(\epsilon_F) \, . \label{Eq_DefThermal2}
\end{align}
Obviously, the coefficients $\sigma_\mathrm{sE}$ and $P_\mathrm{sE}$, which have already been investigated in detail in the previous section, are the only ingredients necessary to fully determine the thermoelectric linear response in the charge sector.

\subsection{Spin Hall magnetoresistance} \label{Sec_SMR}
The SMR is measured under the condition of a vanishing temperature gradient, $\nabla_x T = 0$. The corresponding resistivity, $\rho$, is defined by
\begin{equation}
E_x = \rho \average{j_x} \, .
\end{equation}
Since we are interested in the dependence on the orientation of the attached ferromagnet, we define the ferromagnetic contribution, in analogy to Eq.\ (\ref{Eq_defFM}), by
\begin{equation}
\Delta \rho = \rho - \rho(\gr=0) \, .
\end{equation}
Using Eq.\ (\ref{Eq_jx2}) and assuming $\Delta \rho \ll \rho(\gr=0)$, we obtain
\begin{equation}
\Delta \rho = - \Delta \sigma\rho^2(\gr=0) \, ,
\end{equation}
where
\begin{equation}
\Delta \sigma = -2 e \frac{\alpha}{\hbar}\frac{\tau}{\lDP} \left( \Delta \sigma_\mathrm{sE} - \frac{\lDP}{\taus} \Delta P_\mathrm{sE} \right)
\end{equation}
is the ferromagnetic contribution to the conductivity. Correspondingly, $\Delta \sigma_\mathrm{sE}$ and $\Delta P_\mathrm{sE}$ are the ferromagnetic contributions to the spin Hall conductivity and the polarization coefficient, respectively. Apparently, both $\Delta \sigma_\mathrm{sE}$ and $\Delta P_\mathrm{sE}$ contribute linearly to $\Delta \rho$, and thus the notion ``spin Hall'' magnetoresistance might be misleading in a Rashba system as the one we consider. Yet, since it is extremely difficult to distinguish between the spin Hall and the inverse spin galvanic contributions in an experiment, we stick to this terminology.

\begin{figure}[tb]
\subfigure[]{\includegraphics[width=0.49\columnwidth]{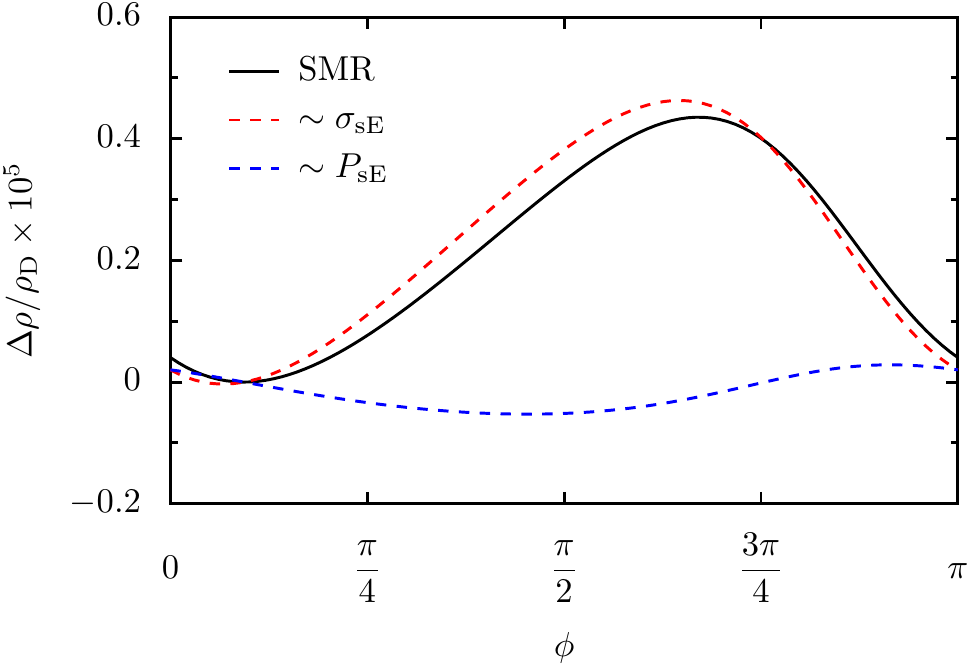}}
\subfigure[]{\includegraphics[width=0.49\columnwidth]{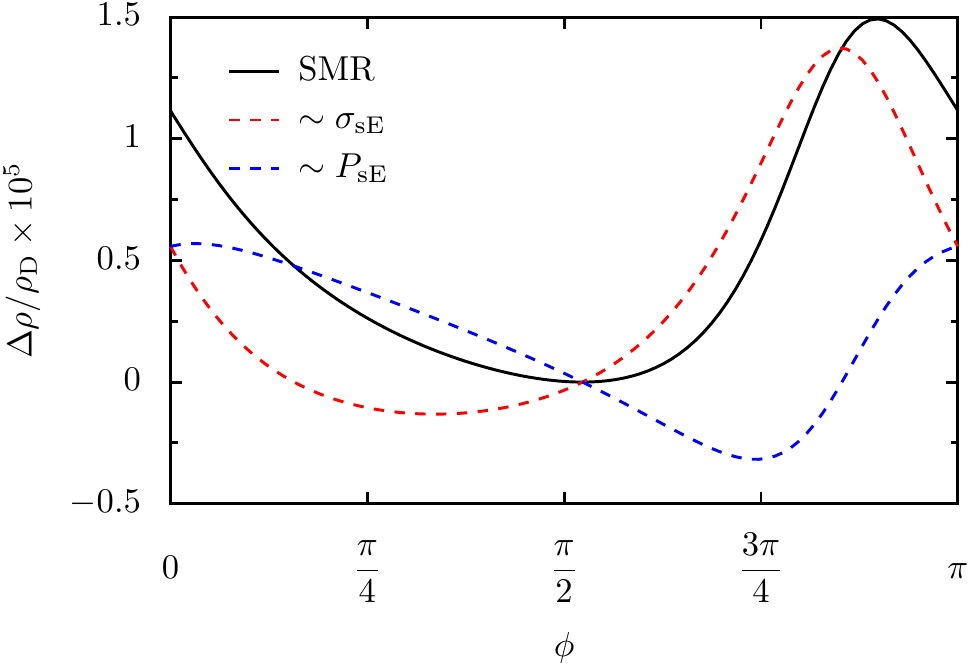}}
\caption{Ferromagnetic contribution to the SMR as function of $\phi$ with $\taus/\tauDP = 10$ and $\gr \alpha \tauDP /\hbar = 10$ for $L = 10\,\lDP$, (a), and $L = \lDP$, (b). The dashed curves represent the contributions proportional to $\sigma_\mathrm{sE}$ (red) and $P_\mathrm{sE}$ (blue), respectively. All data are normalized by $\rho_\mathrm{D}=1/\sigmaD$. \label{Fig_sigmaD}}
\end{figure}

Figure \ref{Fig_sigmaD} shows $\Delta \rho$ versus the magnetization angle $\phi$. For a wide system, (a), the SMR is dominated by the spin Hall ($\sigma_\mathrm{sE}$) contribution, whereas for a narrow system, (b), both contributions appear equally important. Interestingly, at the universal crossing point $\phi_0$ that has already been discussed in the previous section, the contributions $\sim\Delta \sigma_\mathrm{sE}$ and $\sim \Delta P_\mathrm{sE}$ cancel up to linear order such that $\Delta \rho$ has a local minimum at $\phi_0$. In the limit $L \gg \lDP$ it is straightforward to verify this cancellation analytically. Since the ratio $\taus/\tauDP$ can be calculated once $\phi_0$ is known, see Eq.\ (\ref{Eq_phi0}), it is, in principle, possible to extract this ratio experimentally by measuring $\phi_0$.

\subsection{Spin Nernst magnetothermopower}\label{Sec_Seebeck}
Now, we consider a thermal gradient in $x$ direction and study the SNMTP under an open circuit condition, i.e., $\average{j_x}=0$. The thermopower, $S$, is defined by
\begin{equation}
E_x = S \nabla_x T \, .
\end{equation}
Using Eqs.\ (\ref{Eq_jx2})--(\ref{Eq_sygeneral}) we obtain
\begin{equation}
S = \rho \sigmaD \left[1 + 2 \frac{\alpha\tau}{\hbar\lDP} \frac{e}{S_0 \sigmaD} \left( \sigma_\mathrm{ sT} - \frac{\lDP}{\taus} P_\mathrm{ sT} \right) \right] S_0 \, ,
\end{equation}
where 
\begin{equation}
\rho = \frac{1}{\sigmaD} \left[ 1- 2\frac{\alpha\tau}{\hbar\lDP} \frac{e}{\sigmaD} \left( \sigma_\mathrm{sE} - \frac{\lDP}{\taus} P_\mathrm{sE} \right)\right]^{-1}
\end{equation}
is the resistivity corresponding to the SMR as discussed in Sec.\ \ref{Sec_SMR}.
In analogy to Eq.\ (\ref{Eq_defFM}), we define the ferromagnetic contribution to the thermopower by
\begin{equation}
\Delta S = S - S(\gr = 0) \, .
\end{equation}
Keeping only terms linear in $\sigma_\mathrm{sE}$ and $P_\mathrm{sE}$, respectively, it is possible and convenient to split $\Delta S$ into two parts, an electrical part, associated with $\sigma_\mathrm{sE}$ and $P_\mathrm{sE}$, and a thermal part, associated with $\sigma_\mathrm{sT}$ and $P_\mathrm{sT}$. We obtain
\begin{equation}
\Delta S = \Delta S_\mathrm{sE} + \Delta S_\mathrm{sT} 
\end{equation}
with the electrical and thermal parts given by
\begin{align}
\Delta S_\mathrm{sE} &= \Delta \rho\sigmaD S_0  \, , \\
\Delta S_\mathrm{sT} &= 2  e \frac{\alpha}{\hbar}\frac{\tau}{\lDP} \left(  \Delta \sigma_\mathrm{sT} - \frac{\lDP}{\taus} \Delta P_\mathrm{sT} \right) \rho(\gr=0) \, ,
\end{align}
where $\Delta \sigma_\mathrm{sT}$ and $\Delta P_\mathrm{sT}$ are the corresponding ferromagnetic contributions to the direct spin Nernst conductivity and the direct thermal polarization coefficient, respectively.

\begin{figure}[tb]
\subfigure[]{\includegraphics[width=0.49\columnwidth]{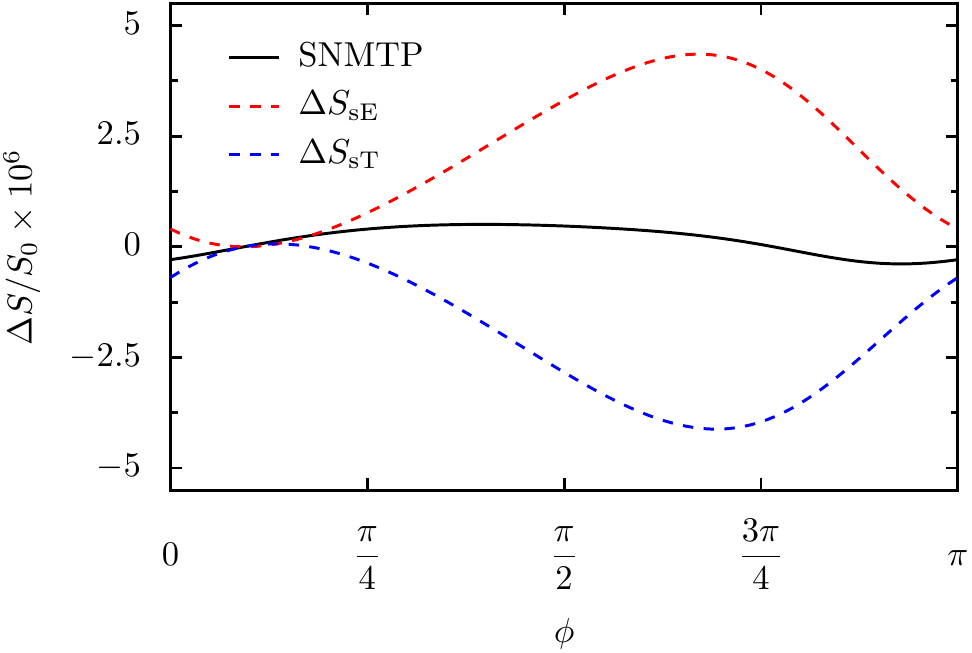}} 
\subfigure[]{\includegraphics[width=0.49\columnwidth]{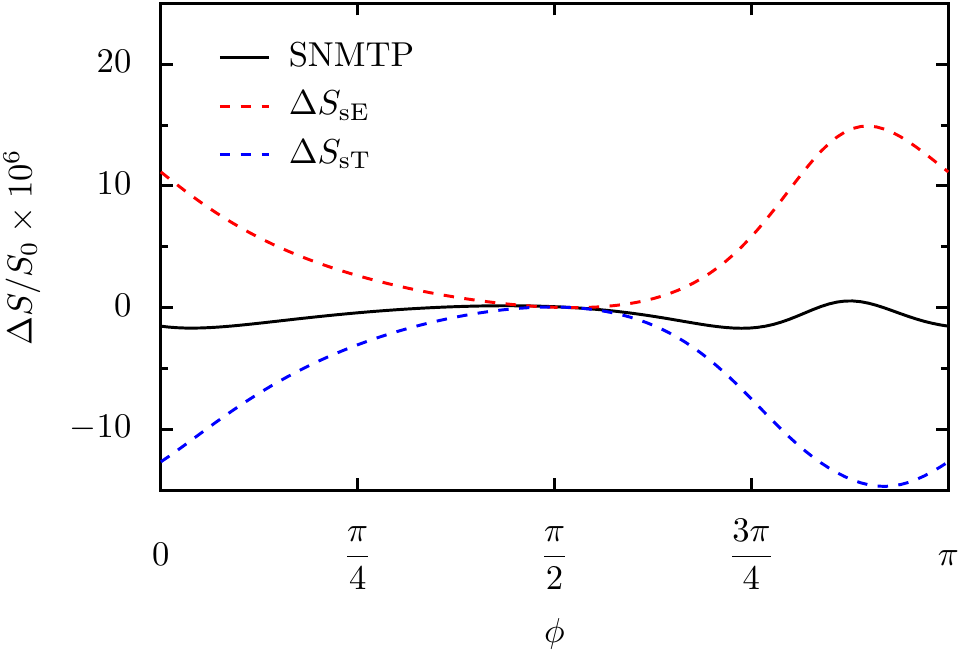}} 
\caption{Ferromagnetic contribution to the SNMTP as function of $\phi$ with $\taus/\tauDP = 10$ and $\gr \alpha \tauDP /\hbar = 10$ for $L = 10\,\lDP$, (a), and $L = \lDP$, (b). The dashed curves represent the electrical part (red) and the thermal part (blue), respectively. \label{Fig_Seebeck}}
\end{figure}

Figure \ref{Fig_Seebeck} shows the SNMTP and its respective electrical and thermal parts as function of the magnetization angle $\phi$. 
Interestingly, electrical and thermal contributions nearly cancel
each other resulting in a rather small SNMTP fingerprint in the thermopower 
for both a wide, (a), and a narrow, (b), system. For the parameters considered
in Fig.~\ref{Fig_Seebeck} this results in $\Delta S/S_0$ being of the order of $10^{-6}$.
Moreover, it can be shown that in the limit of infinitely large 
spin mixing conductance $\gr \to \infty$, and for $\tauDP / \taus \to 0$, this
cancellation is exact such that the SNMTP is completely absent in this case.

\section{Conclusions}\label{Sec_conclusion}
To summarize, we have investigated the spin and charge dynamics of a 
two-dimensional electron gas with Rashba spin-orbit coupling and Elliott-Yafet spin relaxation. 
In particular, we have focused on two recently discussed effects, namely the spin Hall magnetoresistance 
and the spin Nernst magnetothermopower. Based on a generalized Boltzmann equation we have derived a set of 
coupled spin diffusion equations and solved them for boundary conditions that reflect the presence of a
ferromagnetic insulator attached to the two-dimensional electron gas. The two main effects associated with 
spin-orbit coupling, the spin Hall effect and the inverse spin galvanic effect, are significantly affected by
the polarization direction of the ferromagnet due to the spin transfer torque across the interface.
Interestingly, there is a particular polarization direction where both effects are independent of the
spin mixing conductance, which in turn leads to a local minimum in the spin Hall magnetoresistance signature.
The spin Nernst magnetothermopower turns out to be very small due to
a cancellation of electrical and thermal contributions, and it vanishes completely
in the limit of infinite spin mixing conductance if Elliott-Yafet spin relaxation 
is neglected.
Our findings deviate substantially from the results of previous theoretical considerations based on phenomenological drift-diffusion equations.
However, quantitative comparison of our results with published experimental investigations of heavy-metal/magnetic-insulator bilayers, e.g., Pt/YIG, are hardly possible due to different geometries 
and the lack of an accepted microscopic model of the spin-orbit coupling in these metals. It would therefore
be interesting to measure the spin Hall magnetoresistance and the spin Nernst magnetothermopower in
semiconductor heterostructures with pure Rashba spin-orbit coupling, such as suggested in this paper.

\begin{acknowledgments}
We acknowledge stimulating discussions with C.\ Back and L.\ Chen, 
as well as financial support from the German Research Foundation (DFG) through TRR 80
and SFB 689.
\end{acknowledgments}

\appendix
\section{Derivation and general solution of the spin diffusion equations}
\label{App_spin_sector}
The spin sector of the (static) Boltzmann equation is given by the trace of the Boltzmann equation multiplied with $\bsigma$, and can be written as
\begin{equation}
\mathbb{M} \bff = \mathbb{N} \langle \bff \rangle + \mathbf{S} \, , \label{Eq_SpinSector}
\end{equation}
with 
\begin{align}
\mathbb{M} &= 
2- \mathbb{N}
+ \frac{\tau p_y}{m} \nabla_y
 + \frac{2 \alpha \tau}{\hbar^2} 
\begin{pmatrix}
0 & 0 & p_x \\
0 & 0 & p_y \\
-p_x & -p_y & 0 
\end{pmatrix} \, , \\
\mathbb{N} &= 1 - \frac{\tau}{2\taus} 
\begin{pmatrix}
1 & 0 & 0 \\
0 & 1 & 0 \\
0 & 0 & 0 
\end{pmatrix} \, , \\
\mathbf{S} &= \frac{\tau B_z^z}{2m}  \left( \bp \times \bz \right) \cdot \left( \nabla_\bp f^0 \right) \bz + \frac{1}{N_0} \! \left( \frac{\lambda}{2\hbar} \right)^4 \!\!\!\int \frac{\rd^2 p'}{(2\pi\hbar)^2} \bcalA_i L_i \! \left( f^0_\bp - f^0_{\bp'} \right) \! \delta (\epsilon -\epsilon')  ,
\end{align}
where $L_i = ({p'}^2+\bp\cdot\bp')p_i - (p^2+\bp\cdot\bp')p'_i$.
An integration over the momentum and using $j_x = \sigmaD E_x$ leads to the following equations for the $y$- and $z$-component:
\begin{align}
s^y &= -  \taus\nabla_y j_y^y - \frac{\taus}{\lDP}j_y^z  + \frac{\hbar\sigmaD}{4 e \epsilon_F \lDP}  E_x \, , \label{Eq_AppsyDGL} \\
\nabla_y j_y^z &= \frac{1}{\lDP}\left(j_x^x + j_y^y\right) \, , \label{Eq_sz0}
\end{align}
where Eq.\ (\ref{Eq_AppsyDGL}) coincides with Eq.\ (\ref{Eq_syDGL}) in Sec.\ \ref{Sec_SHE_EE}.
Furthermore, we rewrite Eq.~(\ref{Eq_SpinSector}) as
\begin{equation} \label{Eq_SpinSector2}
\bff = \mathbb{M}^{-1}\left( N \langle \bff \rangle + \mathbf{S} \right) \, ,
\end{equation}
where, in the diffusive limit and with $\taus \gg \tau$,
\begin{equation}
\mathbb{M}^{-1} \approx 1- \frac{\tau p_y}{m} \nabla_y  - \frac{2 \alpha \tau}{\hbar^2} 
\begin{pmatrix}
0 & 0 & p_x \\
0 & 0 & p_y \\
-p_x & -p_y & 0 
\end{pmatrix} \, .
\end{equation}
By multiplying Eq.~(\ref{Eq_SpinSector2}) with $p_{x,y}/m$ and integrating over the momentum, we get
\begin{align}
j_x^x &= - \frac{D s^z}{\lDP} \, , \label{Eq_jxx} \\
j_y^y &= -D \nabla_y s^y - \frac{D s^z}{\lDP} \, , \label{Eq_jyy00} \\
j_y^z &= -D \nabla_y s^z + \frac{D s^y}{\lDP} + \frac{\hbar\sigmaD}{4e\epsilon_F \tauDP}  E_x \, . \label{Eq_jyz00}
\end{align}
Inserting Eq.~(\ref{Eq_jxx}) into Eq.~(\ref{Eq_sz0}) gives
\begin{equation}
s^z = -\tauDP \nabla_y j_y^z + \frac{\tauDP}{\lDP} j_y^y \, , \label{Eq_AppszDGL}
\end{equation}
as presented by Eq.\ (\ref{Eq_szDGL}) in the main text. We insert Eqs.~(\ref{Eq_AppsyDGL}) and (\ref{Eq_AppszDGL}) into Eqs.~(\ref{Eq_jyy00}) and (\ref{Eq_jyz00}), respectively, and obtain the following coupled differential equations:
\begin{align}
\left( 2- \ls^2 \nabla_y^2 \right) j_y^y &= \frac{\ls^2+\lDP^2}{\lDP} \nabla_y j_y^z \, ,\\
\left( 1+\frac{\taus}{\tauDP}- \lDP^2 \nabla_y^2 \right) j_y^z &= - \frac{\ls^2+\lDP^2}{\lDP} \nabla_y j_y^y + \frac{\hbar\sigmaD}{2e \epsilon_F \tauDP}  E_x \, ,
\end{align}
cf.\ Eqs.\ (\ref{Eq_jyyDGL}) and (\ref{Eq_jyzDGL}) in Sec.\ \ref{Sec_SHE_EE}.
The general solution of the latter set of equations is given by\footnote{The solutions presented here are valid for $\taus \gg \tauDP$. More generally, these solutions are still correct when the requirement $\taus/\tauDP > 1/(5+4\sqrt{2})$ is fulfilled such that $q_{+}$ is real.}
\begin{align}
j_y^y =& \re^{q_{-} y} \! \left[ ( A_{-} + B_{+}) \cos(q_{+} y) - (A_{+} - B_{-})\sin(q_{+} y) \right]  \nonumber \\
& -  \re^{-q_{-} y} \! \left[ (C_{-} - D_{+}) \cos(q_{+} y)  +  (C_{+} + D_{-})\sin(q_{+} y) \right] , \label{Eq_jyy_general}\\
j_y^z =& j_0^z + \re^{q_{-} y} \! \left[ A \cos(q_{+} y) + B\sin(q_{+} y) \right] +  \re^{-q_{-} y} \! \left[ C \cos(q_{+} y) + D\sin(q_{+} y) \right]  \, , \label{Eq_jyz_general}
\end{align}
where $q_\pm$ is given in Eq.~(\ref{Eq_q}), and
\begin{align}
A_\pm &= \frac{\tauDP}{\tauDP + \taus}\frac{q_\pm}{2} \left(2 \pm \ls^2 |q|^2\right) A \, ,
\end{align}
with $|q|^2 \equiv q_{+}^2+q_{-}^2$; $B_\pm$, $C_\pm$, and $D_\pm$ are defined analogously to $A_\pm$.

\section{Large system sizes}
\label{App_semi}
For $L \gg \lDP$ it is sufficient to consider a semi-infinite system with appropriate boundary conditions at $y=0$, and construct the
approximate solution for finite systems by applying the symmetry relations discussed in the main text, see Sec.\ \ref{Sec_spatial_profile}. 

For $\gr = 0$ the spin currents must vanish at the interface, and the boundary conditions read
\begin{align}
j_y^y(0) &= 0 \, , \,\,\, j_y^y(y\rightarrow \infty) = 0 \, , \\
j_y^z(0) &= 0 \, , \,\,\, j_y^z(y\rightarrow \infty) = j_0^z \, .
\end{align}
Adjusting the general solution of Eqs.~(\ref{Eq_jyy_general}) and (\ref{Eq_jyz_general}) to these boundary conditions yields the spin currents
\begin{align}
j_y^y &= \frac{j_0^z}{2+\ls^2|q|^2} \frac{\lDP |q|^2}{q_{+}} \left( 1+ \frac{\taus}{\tauDP} \right)  \re^{-q_{-} y} \sin(q_{+} y) \, , \\
j_y^z &= j_0^z - \frac{j_0^z }{2+\ls^2|q|^2} \re^{-q_{-} y} \bigg[\!\! \left( 2 + \ls^2|q|^2 \right) \! \cos(q_{+} y) \! + 
\! \frac{q_{-}}{q_{+}} \! \left( 2 - \ls^2|q|^2 \right) \! \sin(q_{+} y) \bigg] \,  .
\end{align}
Using Eqs.~(\ref{Eq_AppsyDGL}) and (\ref{Eq_AppszDGL}) we find the corresponding expressions for the spin densities,
\begin{align}
s^y &= s_0^y + \frac{2 s_0^y}{2+\ls^2|q|^2} \frac{\taus}{\tauDP-\taus} \re^{-q_{-} y} \bigg[\!\! \left( 2 - \lDP^2|q|^2 \right)\! \cos(q_{+} y) \! + \! \frac{q_{-}}{q_{+}} \! 
\left( 2 + \lDP^2|q|^2 \right) \! \sin(q_{+} y)  \bigg] \, , \label{Eq_sy_App_G} \\
s^z &= - \frac{s_0^y}{2+\ls^2|q|^2} \frac{\taus}{\tauDP-\taus} \re^{-q_{-} y}  \bigg[ 4\lDP^3 q_{-}  |q|^2 \cos(q_{+} y) \! + \! \frac{\tauDP-\taus}{\taus} \frac{\lDP|q|^2}{q_{+}} \! \sin(q_{+} y)  \bigg] \, .  \label{Eq_sz_App_G}
\end{align}
For $\gr > 0$ the boundary conditions for a semi-infinite system read
\begin{align}
j_y^y(0) &= j^y_{\mathrm{ FM}} \, , \,\,\, j_y^y(y\rightarrow \infty) = 0 \, , \\
j_y^z(0) &= j^z_{\mathrm{ FM}} \, , \,\,\, j_y^z(y\rightarrow \infty) = j_0^z \, ,
\end{align}
where, for the time being, we assume that the boundary values of the currents, $j^y_{\mathrm{ FM}}$ and $j^z_{\mathrm{ FM}}$, are given.
Matching the general solution, Eqs.~(\ref{Eq_jyy_general}) and (\ref{Eq_jyz_general}), to the boundary conditions we obtain
\begin{align}
\Delta j_y^y &= \frac{\re^{-q_{-} y}}{2+\ls^2|q|^2} \Bigg\{j^y_\mathrm{ FM} \Big[ \left(2+\ls^2|q|^2\right)\cos(q_{+} y) - \frac{q_{-}}{q_{+}} \left(2-\ls^2|q|^2\right) \sin(q_{+} y)  \Big] \nonumber \\
		& \hspace*{2cm}  -j^z_\mathrm{ FM}  \left(1+\frac{\taus}{\tauDP}\right) \frac{\lDP |q|^2}{q_{+}} \sin(q_{+} y) \Bigg\}   \, , \label{Eq_jyyFM} \\
\Delta j_y^z  &= \frac{\re^{-q_{-} y}}{2+\ls^2|q|^2} \Bigg\{j^y_\mathrm{ FM} \left(1+\frac{\taus}{\tauDP}\right) \frac{2}{\lDP q_{+}} \sin(q_{+} y)  \nonumber \\
		& \hspace*{2cm}  +j^z_\mathrm{ FM} \Big[ \left(2+\ls^2|q|^2\right)\cos(q_{+} y) + \frac{q_{-}}{q_{+}} \left(2-\ls^2|q|^2\right) \sin(q_{+} y)  \Big] \Bigg\} \, . \label{Eq_jyzFM}
\end{align}
where $\Delta \bj_y = \bj_y(\gr) - \bj_y(\gr=0)$ is the additional contribution due to the coupling to the ferromagnet.

Let us now consider the boundary values $j^y_{\mathrm{ FM}}$ and $j^z_{\mathrm{ FM}}$ which, according to Eq.\ (\ref{Eq_FM_boundary}), are given by
\begin{equation}\label{Eq_jFM_App}
\bj_\mathrm{FM} = \bj_y (0) = \frac{\gr}{2\pi\hbar N_0} \bn \times \big(\bn \times \bs(0)\big) \, .
\end{equation}
By inserting Eqs.~(\ref{Eq_jyyFM}) and (\ref{Eq_jyzFM}) into Eqs.~(\ref{Eq_AppsyDGL}) and (\ref{Eq_AppszDGL}), we find the ferromagnetic contribution to the spin density which depends through $\bj_\mathrm{FM}$ on the total spin density $\bs(0)$. It is therefore possible to relate $\bs(0)$ to the $\gr=0$ contribution:
\begin{equation} \label{Eq_s_G}
\begin{pmatrix}
s^y(0) \\ s^z(0)
\end{pmatrix} = \left.\begin{pmatrix}
s^y(0) \\ s^z(0)
\end{pmatrix} \right|_{\gr=0} + \mathbb{F} \begin{pmatrix}
s^y(0) \\ s^z(0)
\end{pmatrix} , \,
\end{equation}
where  
\begin{align}
\mathbb{F} =& - \frac{2 \gr \alpha \tauDP}{\hbar} \frac{2\taus/\tauDP-\ls^2|q|^2}{2+\ls^2|q|^2}   \nonumber \\
&\times n_y n_z \begin{pmatrix}
1 + \frac{4 \lDP q_{-}}{2-\lDP^2|q|^2} \frac{n_z}{n_y} \, &	\,	- \frac{n_y}{n_z} - \frac{4 \lDP q_{-}}{2-\lDP^2|q|^2}   \\
- \frac{n_z}{n_y} - \frac{2 \lDP^3 |q|^2 q_{-}}{2-\lDP^2|q|^2}  \,& \,	1	+ \frac{2 \lDP^3 |q|^2 q_{-}}{2-\lDP^2|q|^2} \frac{n_y}{n_z}
\end{pmatrix}  
\end{align}
captures the influence of the ferromagnetic boundary.
Solving Eq.\ (\ref{Eq_s_G}) for $\bs(0)$ yields
\begin{equation} \label{Eq_s0APP}
\begin{pmatrix}
s^y(0) \\ s^z(0)
\end{pmatrix} = \left(1- \mathbb{F} \right)^{-1}\left.\begin{pmatrix}
s^y(0) \\ s^z(0)
\end{pmatrix} \right|_{\gr=0} \, .
\end{equation}
It is convenient to rewrite the inverse matrix in the form
\begin{equation} \label{Eq_1-F}
\left(1-\mathbb{F}\right)^{-1} = \frac{1}{d}\left(1 + \mathbb{G}\right) \, ,
\end{equation}
where 
\begin{align}
d =& 1 + \frac{4\gr \alpha \tauDP}{\hbar} \frac{\taus/\tauDP}{2+\ls^2|q|^2} \Big[ \lDP q_{-} \left( 2 n_z^2 + \lDP^2|q|^2 n_y^2 \right) + \left(2-\lDP^2 |q|^2 \right) n_y n_z \Big]
\end{align}
is the determinant of $1-\mathbb{F}$, and
\begin{align}
\mathbb{G} =& \frac{2\gr \alpha \tauDP}{\hbar} \frac{2 \taus/\tauDP -\ls^2|q|^2}{2+\ls^2|q|^2} \nonumber \\
& \times n_y n_z\begin{pmatrix}
 1 + \frac{2 \lDP^3|q|^2 q_{-}}{2-\lDP^2|q|^2} \frac{n_y}{n_z}   	\,& \,		\frac{n_y}{n_z} + \frac{4\lDP q_{-}}{2-\lDP^2|q|^2}   \\
 \frac{n_z}{n_y} + \frac{2 \lDP^3|q|^2 q_{-}}{2-\lDP^2|q|^2}  	\,& \,		1+ \frac{4\lDP q_{-}}{2-\lDP^2|q|^2} \frac{n_z}{n_y} 
\end{pmatrix} \, .
\end{align}
The matrix $\mathbb{G}$ has the remarkable property 
\begin{equation}
\mathbb{G} \left.\begin{pmatrix}
s^y(0) \\ s^z(0)
\end{pmatrix}\right|_{\gr=0} \sim \begin{pmatrix} n_y \\ n_z \end{pmatrix} \, .
\end{equation}
Therefore, inserting Eqs.\ (\ref{Eq_s0APP}) and (\ref{Eq_1-F}) into the boundary condition, Eq.\ (\ref{Eq_jFM_App}), we obtain
\begin{equation}
\bj_\mathrm{FM} = \frac{\gr}{2\pi\hbar d N_0} \bn \times \big(\bn \times \bs(\gr=0, y=0)\big) \, ,
\end{equation}
which means that the spin polarization for $\gr=0$ fixes the boundary condition for the spin current in the case $\gr > 0$.
With this result it is straightforward to determine the magnetization angle $\phi_0$ for which the ferromagnetic boundary condition is equivalent to the $\gr=0$ boundary condition. For $\bs(\gr=0, y=0) \sim \bn$ the spin current at the interface vanishes. Therefore, the tangent of $\phi_0$ is given by the ratio of $s^z(0)$ and $s^y(0)$ for $\gr=0$. Using Eqs.\ (\ref{Eq_sy_App_G}) and (\ref{Eq_sz_App_G}), we obtain the result given in Eq.~(\ref{Eq_phi0}).




\bibliography{spin_Nernst}
\end{document}